\RequirePackage{lineno}
\documentclass[prl,twocolumn,showpacs,amsmath,amssymb]{revtex4-1}
\usepackage{overpic}
\usepackage{graphicx}
\usepackage{epsfig}
\usepackage{dcolumn}
\usepackage{bm}
\usepackage{rotating}
\usepackage{multirow}
\usepackage{booktabs}
\usepackage{appendix}
\usepackage{subfigure}
\usepackage{hhline}
\usepackage{amssymb}
\usepackage{lineno}
\usepackage{hyperref}
\usepackage{mathrsfs}
\usepackage{verbatim}
\usepackage{amsmath}
\usepackage{amssymb}
\usepackage{amsfonts}
\usepackage{amsbsy}
\usepackage{float}
\usepackage{epstopdf}

\hypersetup{colorlinks,linkcolor=blue,anchorcolor=blue,citecolor=blue}
\include{def-com}

\let\oldequation\equation
\let\oldendequation\endequation

\renewenvironment{equation}
  {\linenomathNonumbers\oldequation}
  {\oldendequation\endlinenomath}

\newcommand{\mev}{\mathrm{MeV}}

\newcommand{\gevcc}{\mathrm{GeV}/c^2}

\def\pip{\pi^{+}}
\def\pim{\pi^{-}}
\def\piz{\pi^{0}}
\def\ee{e^{+}e^{-}}

\def \Ks {K_{S}^{0}}

\def \gevcc{\mbox{GeV/$c^2$}}
\def \mev  {\mbox{MeV}}

\def \miss2{M_{\rm miss}^{2}}

\def \romanOne   {\uppercase\expandafter{\romannumeral1}}
\def \romanTwo   {\uppercase\expandafter{\romannumeral2}}
\def \romanThree {\uppercase\expandafter{\romannumeral3}}
\def \romanFour  {\uppercase\expandafter{\romannumeral4}}
\def \romanFive  {\uppercase\expandafter{\romannumeral5}}
\def \romanSix   {\uppercase\expandafter{\romannumeral6}}
\def \romanSeven {\uppercase\expandafter{\romannumeral7}}
\def \romanEight {\uppercase\expandafter{\romannumeral8}}
\def \romanNine {\uppercase\expandafter{\romannumeral9}}

\newcommand{\lamcplamcm}{\Lambda_{c}^{+}\bar{\Lambda}_{c}^{-}}
\newcommand{\lambdacp}{\Lambda_{c}^{+}}
\newcommand{\lambdacm}{\bar{\Lambda}_{c}^{-}}

\newcommand{\sigmode}[1]{
	\ifnum#1=1
	\lambdacp \rightarrow n K_{S}^{0} \pi^{+}
	\else
	\ifnum#1=2
	\lambdacp \rightarrow n K_{S}^{0} K^{+}
	\fi
	\fi
}

\begin{document}

\title{\boldmath First observation of the decay $\Lambda^+_c\to nK^{0}_{S}\pi^+\pi^0$}

\author{
\begin{small}
\begin{center}
M.~Ablikim$^{1}$, M.~N.~Achasov$^{4,b}$, P.~Adlarson$^{75}$, O.~Afedulidis$^{3}$, X.~C.~Ai$^{80}$, R.~Aliberti$^{35}$, A.~Amoroso$^{74A,74C}$, Q.~An$^{71,58}$, Y.~Bai$^{57}$, O.~Bakina$^{36}$
% ...
\\
\vspace{0.2cm}
(BESIII Collaboration)\\
\vspace{0.2cm} {\it
$^{1}$ Institute of High Energy Physics, Beijing 100049, People's Republic of China\\
% ...
$^{3}$ Bochum  Ruhr-University, D-44780 Bochum, Germany\\
$^{4}$ Budker Institute of Nuclear Physics SB RAS (BINP), Novosibirsk 630090, Russia\\
% ...
$^{35}$ Johannes Gutenberg University of Mainz, Johann-Joachim-Becher-Weg 45, D-55099 Mainz, Germany\\
$^{36}$ Joint Institute for Nuclear Research, 141980 Dubna, Moscow region, Russia\\
% ...
$^{57}$ Southeast University, Nanjing 211100, People's Republic of China\\
$^{58}$ State Key Laboratory of Particle Detection and Electronics, Beijing 100049, Hefei 230026, People's Republic of China\\
% ...
$^{71}$ University of Science and Technology of China, Hefei 230026, People's Republic of China\\
% ...
$^{74}$ University of Turin and INFN, (A)University of Turin, I-10125, Turin, Italy; (B)University of Eastern Piedmont, I-15121, Alessandria, Italy; (C)INFN, I-10125, Turin, Italy\\
$^{75}$ Uppsala University, Box 516, SE-75120 Uppsala, Sweden\\
% ...
$^{80}$ Zhengzhou University, Zhengzhou 450001, People's Republic of China\\
\vspace{0.2cm}
% ...
$^{b}$ Also at the Novosibirsk State University, Novosibirsk, 630090, Russia\\
% ...
}
\end{center}
\vspace{0.4cm}
\end{small}
}

\date{\today}

\begin{abstract}
Based on 4.5 fb$^{-1}$ of $\ee$ collision data accumulated at center-of-mass energies between $4599.53\,\mev$ and $4698.82\,\mev$ with the BESIII detector, the decay $\Lambda_{c}^{+}\to nK_{S}^{0}\pi^+\pi^0$ is observed for the first time with a significance of $9.2\sigma$. The branching fraction is measured to be $(0.85\pm0.13\pm0.03)\%$, where the first uncertainty is statistical and the second systematic, which differs from the theoretical prediction based on isospin by 4.4$\sigma$. This indicates that there may be resonant contributions or some unknown dynamics in this decay.
 
\end{abstract}

\maketitle
\oddsidemargin -0.2cm
\evensidemargin -0.2cm

\section{\boldmath Introduction}

Experimental studies of charmed baryon decays provide important information on the strong and weak interactions in the environment of heavy quarks. The lightest charmed baryon, $\Lambda_c^+$, was first observed in $e^+e^-$ annihilation at the Mark II experiment~\cite{Abrams:1979iu}.
So far, 70\% of the $\Lambda_c^+$ decays have been observed~\cite{ParticleDataGroup:2020ssz,Li:2021iwf}, of which 
the total branching fraction (BF) of the decays involving a neutron is 
$(32.4\pm1.7)\%$~\cite{BESIII:liqingzhang_work}.
To deeply understand the properties of non-perturbative quantum chromodynamics, studies of multi-body hadronic decays which include potential intermediate processes are needed.
Recently, the BFs of the Cabibbo-favored four body decays $\Lambda_c^+\to pK^-\pi^+\pi^0$~\cite{BESIII:2015bjk}, $\Lambda_c^+\to nK^-\pi^+\pi^+$~\cite{BESIII:yinghao_work}, and $\Lambda_c^+\to pK_{S}^{0}\pi^+\pi^-$~\cite{BESIII:2015bjk} were measured to be $(4.53 \pm 0.38)\%$, $(1.90 \pm 0.12)\%$, and $(1.53 \pm 0.14)\%$, respectively. However, the isospin-related decay $\Lambda_c^+\to nK_{S}^{0}\pi^+\pi^0$ has not been observed yet.

\begin{figure}[tbp]\centering

	\hspace{-3mm}
	\subfigure[]{
		\label{fig:nkspiFeynmanA}
		\includegraphics[width=0.2\textwidth]{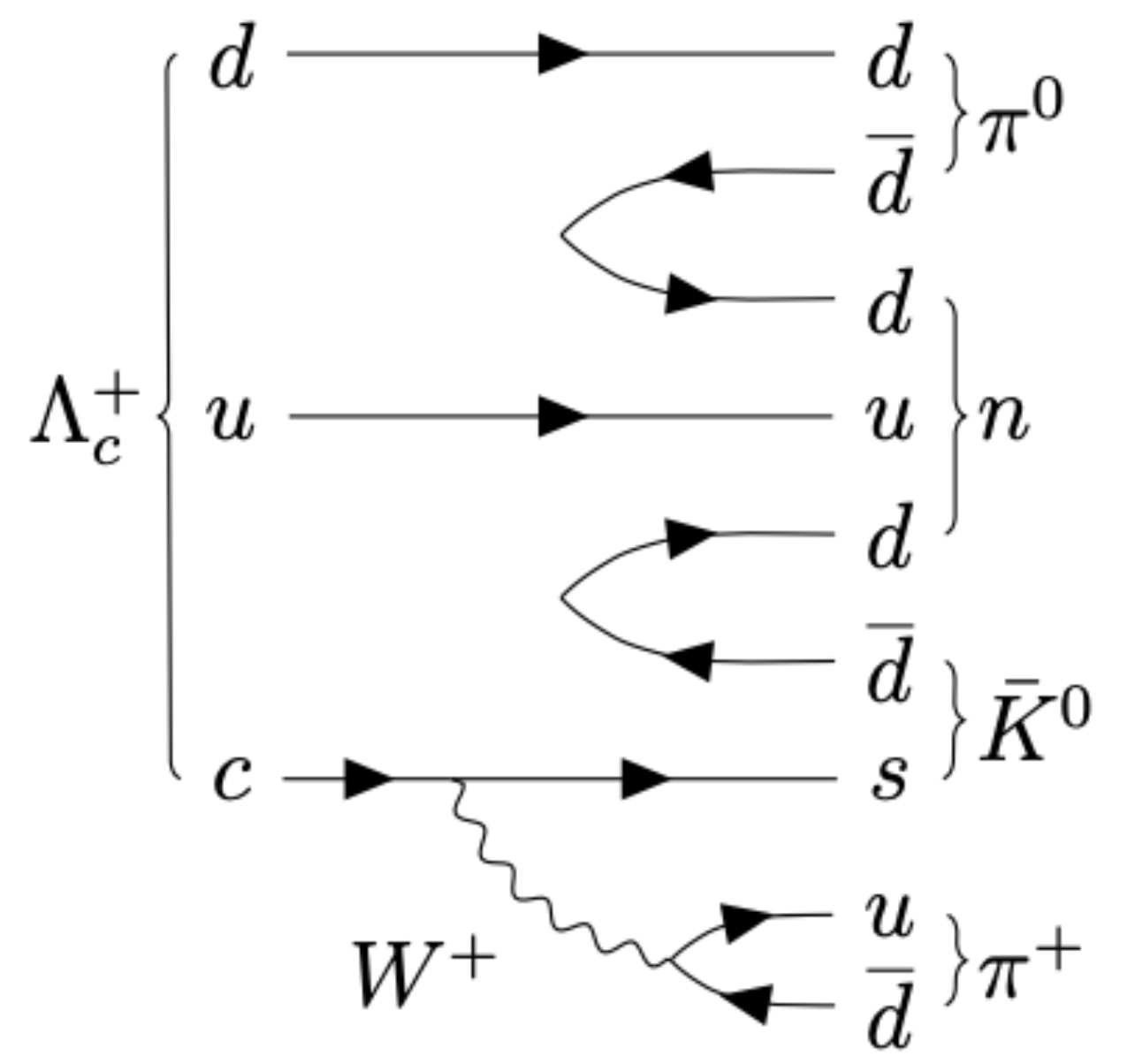}
	}
	\hspace{0mm}
	\subfigure[]{
		\label{fig:nkspiFeynmanB}
		\includegraphics[width=0.2\textwidth]
        {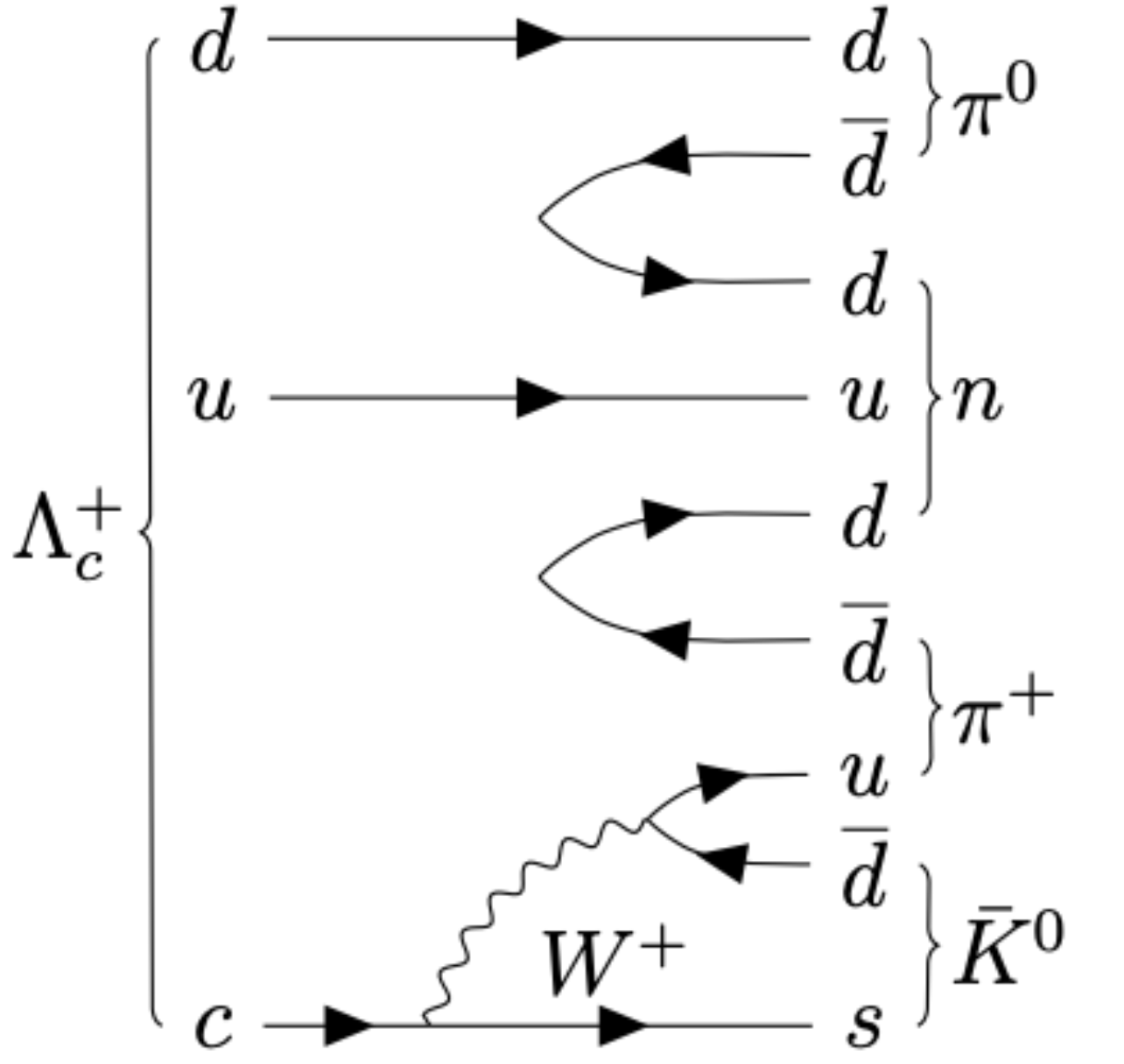}
	}
	\hspace{-15mm}
	\subfigure[]{
		\label{fig:nkspiFeynmanC}
		\includegraphics[width=0.2\textwidth]{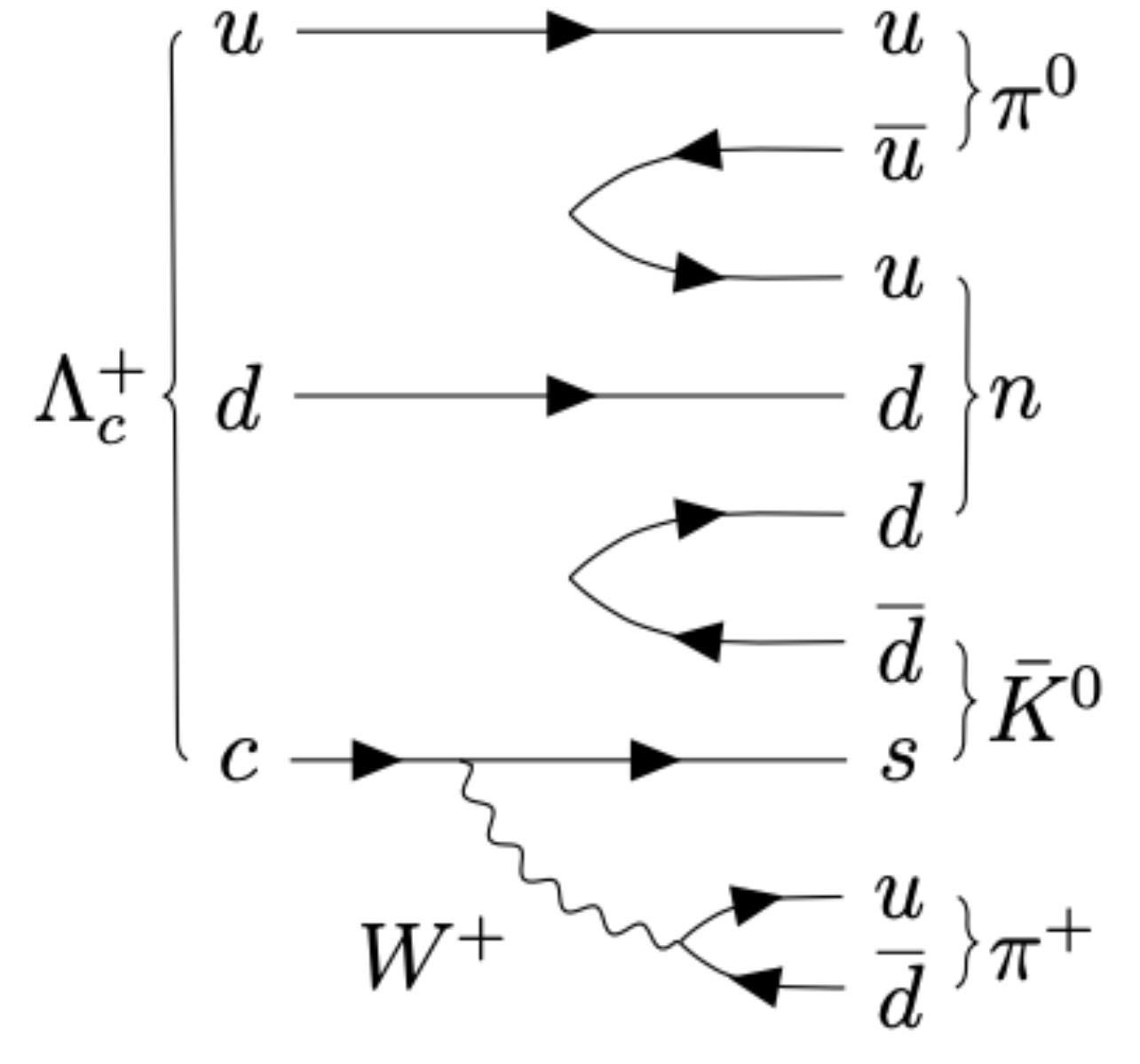}
	}
	\subfigure[]{
		\label{fig:nkspiFeynmanD}
		\includegraphics[width=0.2\textwidth]
        {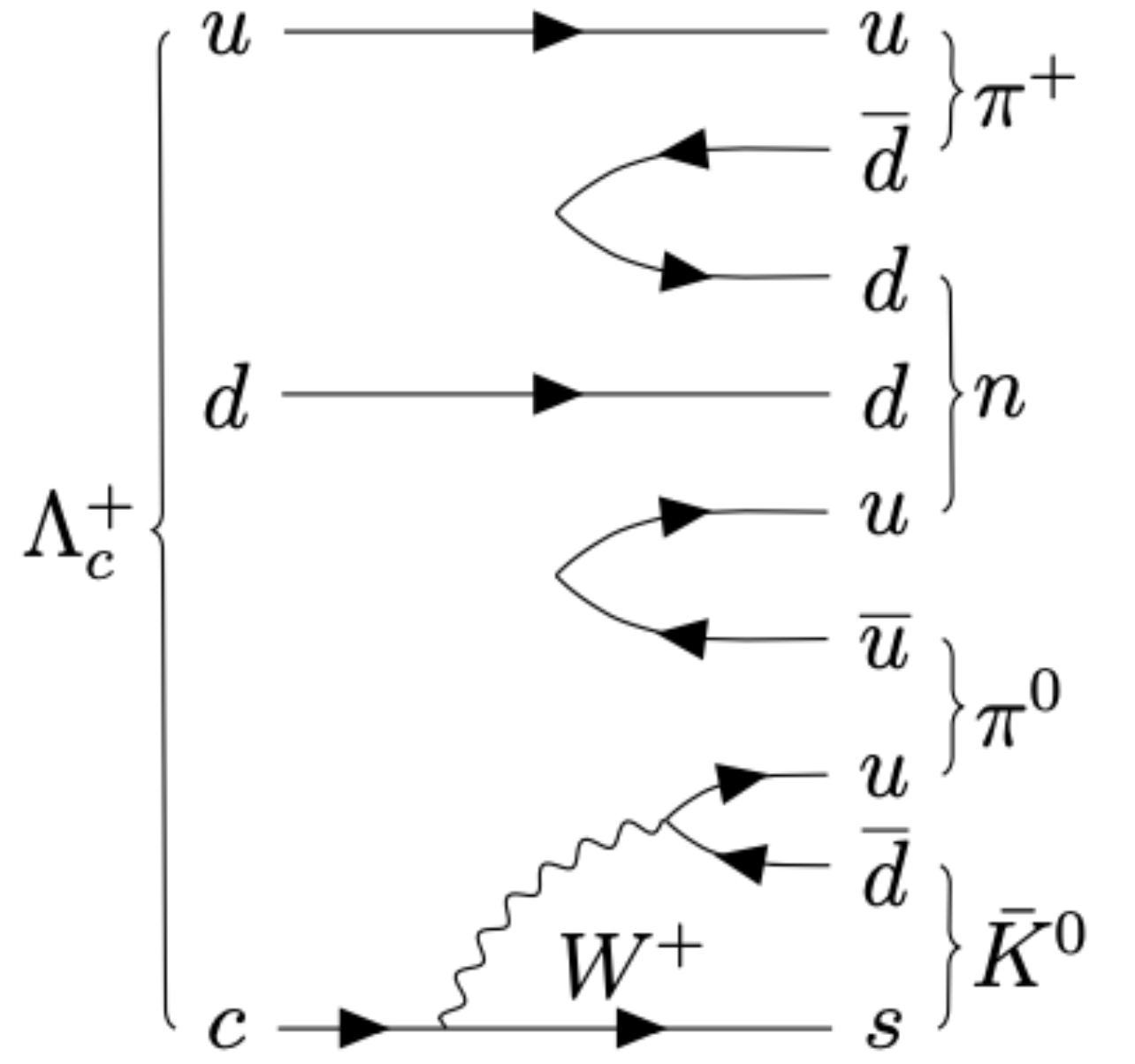}
	}
	\hspace{-15mm}
	\subfigure[]{
		\label{fig:nkspiFeynmanE}
		\includegraphics[width=0.2\textwidth]{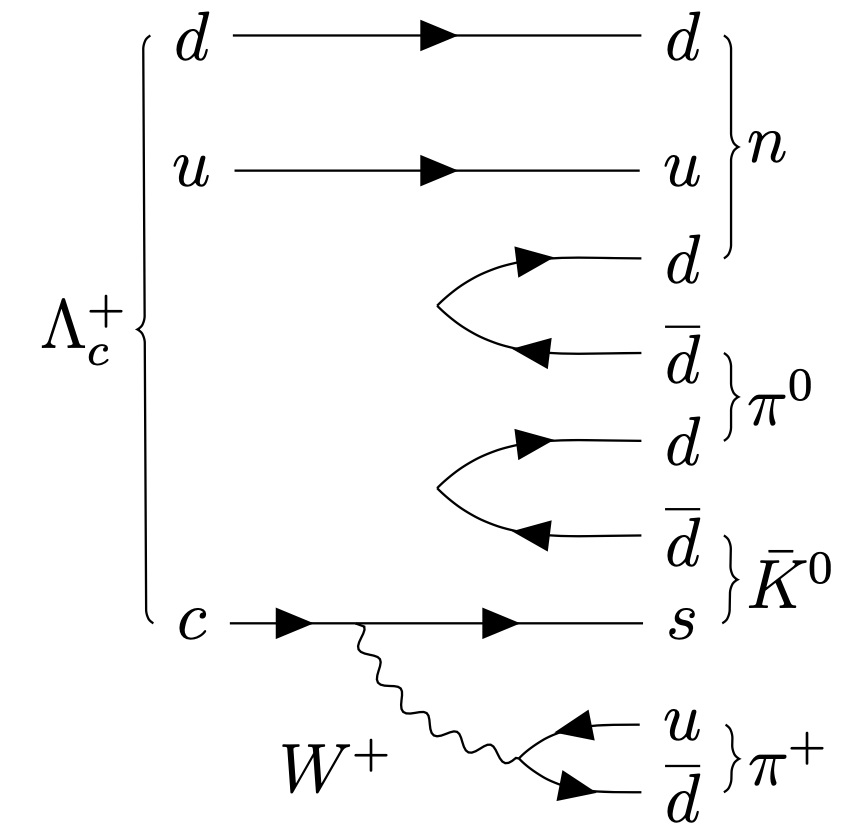}
	}
	\subfigure[]{
		\label{fig:nkspiFeynmanF}
		\includegraphics[width=0.2\textwidth]
        {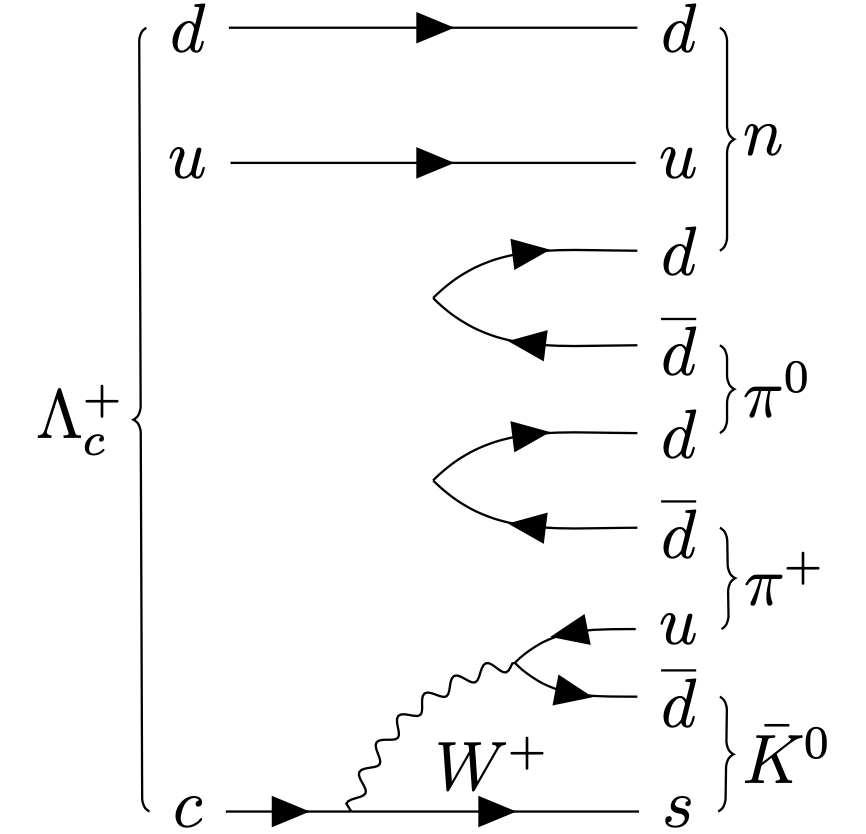}
	}

	\caption{Feynman diagrams for $\Lambda_{c}^{+}\to nK_{S}^{0}\pi^+\pi^0$\,: (a),(c),(e) External $W$-emission; (b),(d),(f) Internal $W$-emission.}
	\label{fig:Feynman}
\end{figure}

Theoretically, the decay amplitude of $\Lambda^+_c$ consists of factorizable and non-factorizable contributions~\cite{Chau:1986jb, Chau:1995gk}. It is known that the non-factorizable contribution is negligible compared to the factorizable one in describing the non-leptonic weak decays of charmed mesons~\cite{Bauer:1986bm}. However, the contributions are different in $\Lambda^+_c$ decays, where the $W$-exchange diagram manifesting a pole diagram is no longer subject to helicity and color suppression~\cite{Cheng:1993gf}. As shown in Fig.~\ref{fig:Feynman}, the $\Lambda_c^+\to nK_{S}^{0}\pi^+\pi^0$ decay proceeds through external and internal $W$-emission processes, where the dynamics includes both factorizable and non-factorizable contributions.
There has been much progress in the theoretical and experimental studies of the two-body decays of the $\lambdacp$~\cite{CHENG2022324}. 
However, due to possible intermediate resonances, the dynamics of multi-body decays of $\Lambda^+_c$ is more complex, and theoretical calculations are not reliable yet. 
The decay $\Lambda_c^+\to nK_{S}^{0}\pi^+\pi^0$ is dominated by the weak interaction process $c\to s u\bar{d}$.
A phenomenological model based on isospin~\cite{Gronau:2018vei} predicts a BF for this decay of $(1.54\pm0.08)\%$ and further relates 
the BFs for all $N\bar{K}\pi\pi$ final states.  

In this paper, we report the first observation of $\Lambda_c^+\to nK_{S}^{0}\pi^+\pi^0$ based on the data samples accumulated at center-of-mass (c.m.) energies between $4599.53\,\mev$ and $4698.82\,\mev$ with the BESIII detector. These data samples correspond to an integrated luminosity of $4.5~\rm fb^{-1}$~\cite{BESIII:2015zbz,BESIII:2022dxl,BESIII:2022ulv,Ke:2023qzc}, as listed in Table~\ref{tot_lum1}. 
Due to the energy being just above the threshold of $\lambdacp\lambdacm$ pair production,
the $\lambdacp\lambdacm$ pairs are generated without other accompanied hadrons, which makes it feasible to apply the double-tag (DT) method~\cite{MARK-III:1985hbd} and reconstruct the neutron with missing neutron mass technique.
The $\lambdacp$ is reconstructed by recoiling against the single-tag (ST) candidate $\lambdacm$ at these c.m.~energies, and an event containing an ST $\bar{\Lambda}^-_c$ and a signal $\Lambda^+_c$ is referred to as a DT candidate.  Charge-conjugated decays are implied throughout this paper.
 \begin{table}[htbp]
 \centering

\caption{The c.m.~energies and integrated luminosities for each data sample, where the first uncertainties are statistical, and the second ones are systematic.} 
\label{tot_lum1}
\begin{tabular}{cc}
\hline \hline 
$\sqrt{s}$ (MeV) &Luminosity ($\rm{pb}^{-1}$)\\\hline 
$4599.53\pm0.07\pm0.74$&$586.90\pm0.10\pm3.90$  \\
$4611.86\pm0.12\pm0.32$&$103.65\pm0.05\pm0.55$  \\
$4628.00\pm0.06\pm0.32$&$521.53\pm0.11\pm2.76 $ \\
$4640.91\pm0.06\pm0.38$&$551.65\pm0.12\pm2.92 $ \\
$4661.24\pm0.06\pm0.29$&$529.43\pm0.12\pm2.81 $ \\
$4681.92\pm0.08\pm0.29$&$1667.39\pm0.21\pm8.84 $ \\
$4698.82\pm0.10\pm0.39$&$535.54\pm0.12\pm2.84 $ \\
 \hline \hline
\end{tabular}
\end{table}

\section{\boldmath BESIII Experiment and Monte Carlo Simulation}

The BESIII detector~\cite{Ablikim:2009aa} records symmetric $e^+e^-$ collisions 
provided by the BEPCII storage ring~\cite{Yu:2016cof}
in the c.m.~energy ranging from 2.0 to 4.95~GeV, with a peak luminosity of $1\times10^{33}$~cm$^{-2}$s$^{-1}$ achieved at $\sqrt{s}=3.773$ GeV.
BESIII has collected large data samples in this energy region~\cite{Ablikim:2019hff}. The cylindrical core of the BESIII detector covers 93\% of the full solid angle and consists of a helium-based multilayer drift chamber~(MDC), a plastic scintillator time-of-flight system~(TOF), and a CsI(Tl) electromagnetic calorimeter~(EMC), which are all enclosed in a superconducting solenoidal magnet
providing a 1.0~T magnetic field. The solenoid is supported by an
octagonal flux-return yoke with resistive plate counter muon
identification modules interleaved with steel. 
The charged-particle momentum resolution at $1~{\rm GeV}/c$ is
$0.5\%$, and the resolution of energy deposited (d$E$/d$x$) is $6\%$ for electrons
from Bhabha scattering. The EMC measures photon energies with a
resolution of $2.5\%$ ($5\%$) at $1$~GeV in the barrel (end-cap)
region. The time resolution in the TOF barrel region is 68~ps, while that in the end-cap region is 110~ps. The end cap TOF
system was upgraded in 2015 using multigap resistive plate chamber
technology, providing a time resolution of 60~ps.  About 87\% of the data used in this analysis benefits from this upgrade.  
More detailed descriptions can be found in Refs.~\cite{Ablikim:2009aa, Yu:2016cof}.

Monte-Carlo~(MC) simulated samples are used to determine the detection efficiencies, optimize selection criteria, and study backgrounds. The simulation is carried out with a {\sc geant4}-based~\cite{geant4} package including the geometric description of the BESIII detector and the detector response~\cite{detvis}, and 
models the beam-energy spread and initial-state radiation (ISR) in the $e^+e^-$
annihilation with the generator {\sc kkmc}~\cite{KKMC}. Final-state radiation from charged final-state particles is incorporated using the {\sc photos} package~\cite{photos}. The inclusive MC samples include the production of $\lamcplamcm$ pairs, open-charmed mesons, ISR production of vector charmonium(-like) states, and continuum processes, which are used to determine ST efficiencies, analyze backgrounds, and extract background shapes.  Known decay modes are modeled with {\sc evtgen}~\cite{Lange:2001uf, Ping:2008zz} using BFs taken from the Particle Data Group (PDG)~\cite{ParticleDataGroup:2020ssz}. The remaining unknown charmonium decays are modeled with {\sc lundcharm}~\cite{lundcharm, Yang:2014vra}.
The angular distribution of $e^{+}e^{-}\to\Lambda_{c}^{+}\bar\Lambda_{c}^{-}$ is generated as $1+\alpha\cos^{2}\theta_{\Lambda_{c}^{+}}$, where $\theta_{\Lambda_{c}^{+}}$ is the polar angle between the $\Lambda_{c}^{+}$ and the positron beam in the c.m.~frame, and $\alpha$ is the angular parameter of $\lambdacp$ production, which is different at the seven c.m.~energy points~\cite{BESIII:2023rwv}. 
The signal MC sample consists of the exclusive process where the $\lambdacm$ decays to eleven ST tag modes and the $\lambdacp$ decays to $nK_{S}^{0}\pi^+\pi^0$, with $\Ks\to\pip\pim$ and $\pi^{0}\to\gamma\gamma$, which is used to determine DT efficiencies and extract signal shapes. The $\Lambda_c^+\to nK_{S}^{0}\pi^+\pi^0$ signal MC sample is simulated using a phase space (PHSP) model which then has the four daughter ($K_S^0, n, \pi^+, \pi^0$) momentum and the $n\pi^+\pi^0$ invariant mass distributions weighted to match those of data.

\section{\boldmath Event Selection}
\label{sec:selection}
\begin{figure}[H]
\begin{center}
\begin{overpic}[width=0.48\textwidth,angle=0]{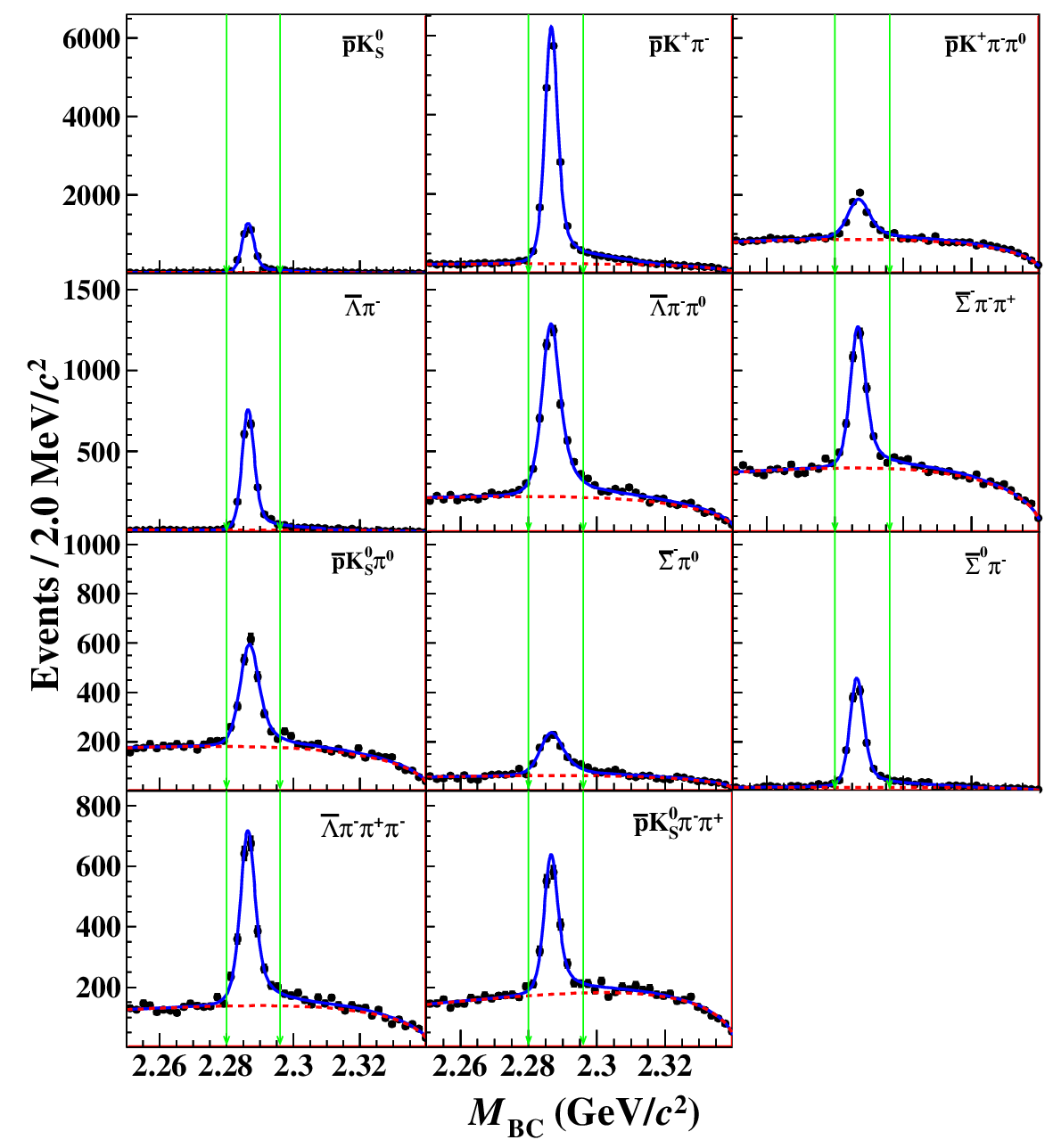}
\end{overpic}
\end{center}
\caption{The $M_{\rm BC}$ distributions of the ST modes for the data sample at $\sqrt{s}=4681.92$ MeV. The points with error bars represent data. The blue solid curves indicate the fit results, the red dashed curves describe the background shapes, and the green lines are defined as signal region.}
\label{fig:mbc}
\end{figure}

The selection criteria of the ST candidates $\bar\Lambda_{c}^{-}\to\bar{p}\Ks$, $\bar{p}K^{+}\pim$,
$\bar{p}K^{+}\pim\piz$,
$\bar{\Lambda}\pim$,
$\bar{\Lambda}\pim\piz$, 
$\bar{\Sigma}^{-}\pim\pip$,
$\bar{p}\Ks\piz$, 
$\bar{\Sigma}^{-}\piz$,
$\bar{\Sigma}^{0}\pim$,  $\bar{\Lambda}\pim\pip\pim$,  and $\bar{p}\Ks\pim\pip$ are the same as Ref.~\cite{BESIII:yinghao_work}. 
The ST $\lambdacm$ is identified with the beam-constrained mass $M_{\rm BC}\equiv\sqrt{E_{\rm beam}^{2}/c^{4}-p_{\Lambda_{c}}^{2}/c^{2}}$, where $E_{\rm beam}$ is the beam energy and $p_{\Lambda_{c}}$ is the measured momentum of the $\lambdacm$ in the c.m.~system of $e^+e^-$ collision. 
 Figure~\ref{fig:mbc} shows the $M_{\rm BC}$ distributions of various ST modes for the data sample at $\sqrt{s}=4681.92$ MeV; clear $\lambdacm$ signals are observed in each mode.
The signal and sideband regions for the ST candidates are defined as $(2.280, 2.296)\,\gevcc$ and $(2.250, 2.270)\,\gevcc$, respectively.
Candidates in the signal region are used for the further DT reconstruction, and those in the sideband region are used to estimate the background contribution.

The $\Ks(\pi^+\pi^-$), $\pi^+$, $\pi^0(\gamma\gamma)$ from the signal side decay 
$\Lambda_c^+\to nK_{S}^{0}\pi^+\pi^0$ are reconstructed with the following criteria.
Charged tracks detected in the MDC are required to be within a polar angle ($\theta$) range of $|\cos\,\theta|<0.93$, where $\theta$ is defined with respect to the $z$-axis which is the symmetry axis of the MDC. 
The $\Ks$ candidate is reconstructed from two oppositely charged tracks satisfying $|V_{z}|<20\,\mathrm{cm}$, where $V_{z}$ denotes the distance to the interaction point (IP) along $z$-axis. 
The two charged tracks are assigned as $\pi^+\pi^-$ without imposing further particle identification~(PID) criteria to improve the $\Ks$ reconstruction efficiency. They are required to originate from a common vertex.
The decay length ($L$) of the $\Ks$ candidate is required to be greater 
than twice the vertex resolution ($\sigma_{L}$) away from IP, $i.e.$, $L/\sigma_{L}>2$. 
If there are multiple $\Ks$ candidates, the one with the largest $L/\sigma_{L}$ is retained. 

Apart from the $\Ks$ candidate, there is one other charged track: the $\pi^+$ from the $\Lambda_c^+$ decay. 
The distance of closest approach to the IP for this track must be less than 10 cm along the z-axis, $|V_{z}|<10\,$cm, and less than 1 cm in the transverse plane, $|V_{xy}|<1\,$cm.
The PID for this track combines the measurements of d$E$/d$x$ in the MDC and the flight time in the TOF to form likelihoods $\mathcal{L}(h)~(h=p,K,\pi)$ for each hadron ($h$) hypothesis.
The track needs to satisfy $\mathcal{L}(\pi)>\mathcal{L}(K)$. 

The photon candidates are identified as showers in the EMC. The deposited energy of each shower must be more than 25 MeV in the barrel region ($|\!\cos \theta|< 0.80$) and more than 50 MeV in the end cap region ($0.86 <|\!\cos \theta|< 0.92$) of the EMC. 
To exclude showers that originate from the charged track radiation, the angle subtended by the EMC shower and the position of the closest charged track at the EMC must be greater than 10 degrees as measured from the IP. To suppress electronic noise and showers unrelated to the event, the difference between the EMC time and the event start time is required to be within [0, 700] ns. 
The invariant mass $m_{\gamma\gamma}$ of the two photons from the $\pi^0$ decay has to satisfy
$0.115 <m_{\gamma\gamma}< 0.150$ GeV$/c^2$. In addition, a one-constraint (1C) kinematic fit is performed to constrain $m_{\gamma\gamma}$ to the $\pi^0$ known mass~\cite{ParticleDataGroup:2020ssz}. 
The fit chi-squared, $\chi^{2}_{1\rm C}$, is required to be less than 200. 
If there are more than one $\pi^0$ candidates, the one with the smallest $\chi^{2}_{1\rm C}$ is retained.

Considering that the neutron is difficult to detect, it is reconstructed with the missing-mass technique, using the kinematic variable $M(n)= \sqrt{E_{\rm miss}^{2}/c^{4}-\,|\vec{p}_{\rm miss}\,|^{2}/c^{2}}$. Here, the $E_{\rm miss}$ and $\vec{p}_{\rm miss}$ are calculated by $E_{\rm miss}= E_{\rm beam}-E_{\rm rec}$ and $\vec{p}_{\rm miss}=\vec{p}_{\lambdacp}-\vec{p}_{\rm rec}$, where $E_{\rm rec}~(\vec{p}_{\rm rec})$ is the sum of the energies (vector momenta) of the reconstructed $\Ks$, $\pi^+$ and $\pi^0$ in the $e^+e^-$ c.m.~system. The $\lambdacp$ momentum $\vec{p}_{\lambdacp}$ is derived by $\vec{p}_{\lambdacp}=-\hat{p}_{\rm tag}\sqrt{E_{\rm beam}^{2}/c^{2}-m_{\lambdacp}^{2}c^{2}}$, where $\hat{p}_{\rm tag}$ is the unit vector of the $\lambdacm$ momentum direction and $m_{\lambdacp}$ is the $\lambdacp$ nominal mass~\cite{ParticleDataGroup:2020ssz}. 

\begin{figure*}[t]

    \begin{overpic}[width=0.9\textwidth]{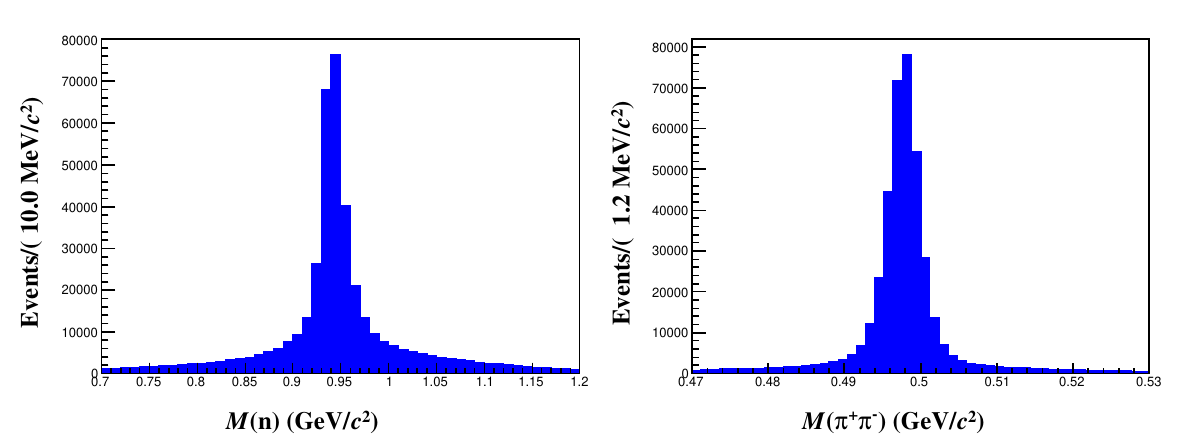}
    \end{overpic}
        \caption{The blue histograms represent $M(n)$ and $M(\pi^+\pi^-)$ distributions of signal MC in the signal region.}
    \label{fig:bkg_dis}
\end{figure*}

A study of the inclusive MC sample shows that the peaking backgrounds are mainly from $\Lambda_{c}^{+} \to \Sigma^-(n\pi^-) \, \pi^+\pi^+\pi^0$, $\Lambda(n\pi^0) \, \pi^+\pi^-\pi^+$, and $\Sigma^+(n\pi^+) \, \omega(\pi^+\pi^-\pi^0)$. In order to remove these peaking backgrounds, 
the difference of the invariant masses $M(n\pim)-M(n)$ is required to be outside the interval 
$(0.22, 0.27)\,\gevcc$ and $M(n\pi^0)-M(n)$ must be larger than $0.20\,\gevcc$. Also, both values of $M(n\pip)-M(n)$ are required to lie outside of 
the interval of $(0.23, 0.28)\,\gevcc$, where the two $M(n\pip)$ combinations include the $\pi^+$ from either the $\Lambda_c^{+}$ or the $K_{S}^0$. 
Here, $M(n\pip)$,  $M(n\pim)$, and $M(n\pi^0)$ denote the invariant masses of the missing neutron and $\pip$, $\pim$, and $\pi^0$, respectively. 
Figure~\ref{fig:bkg_dis} shows $M(n)$ and $M(\pi^+\pi^-)$ distributions of signal MC in the signal region after applying all the above selections.

\section{\boldmath Absolute BF measurement}
\label{sec:bf}

\begin{figure*}[t]

    \begin{overpic}[width=0.9\textwidth]{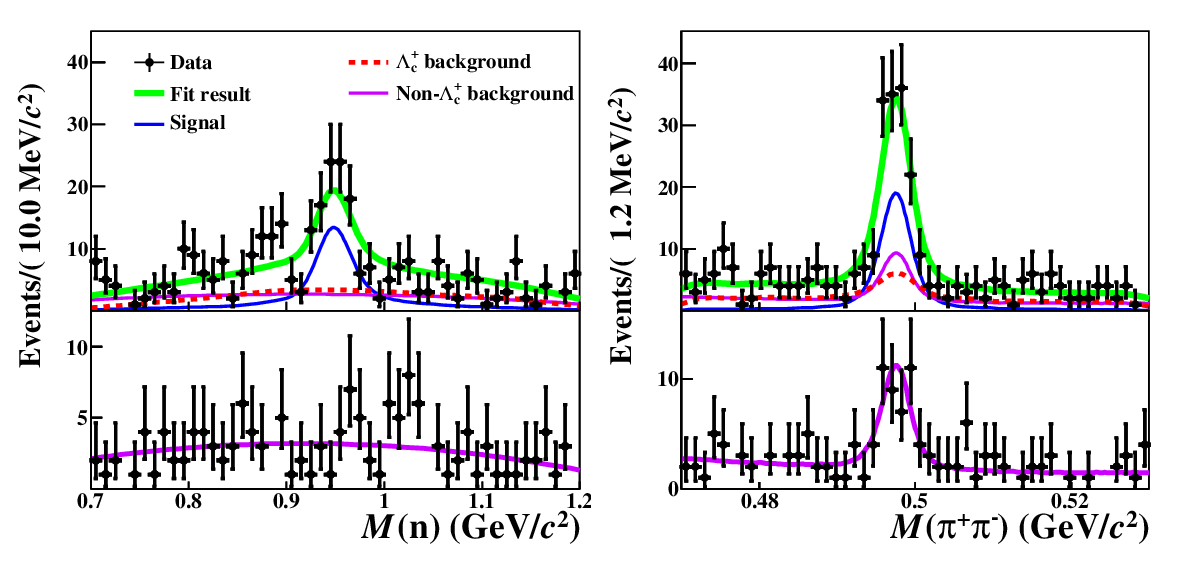}
    \end{overpic}
    \caption{The 2D simultaneous fit on the accepted candidates in the signal (top) and sideband (bottom) regions. The black dots with error bars represent data. The green solid lines represent the total fit results. The blue lines, purple lines, and red dashed lines represent the signal, non-$\Lambda_{c}^{+}\bar{\Lambda}_{c}^{-}$ background, and $\Lambda_{c}^{+}\bar{\Lambda}_{c}^{-}$ pair decay background, respectively.}
    \label{fig:fits}
\end{figure*}

The signal yield of $\Lambda_c^+\to nK_{S}^{0}\pi^+\pi^0$ is obtained by performing a two-dimensional (2D) unbinned maximum likelihood fit to the $M(n)$ and $M(\pip\pim)$ spectra 
of the candidates combined from the data sets at seven c.m.~energies shown 
in Fig.~\ref{fig:fits}. The signal shapes ($f_{\rm sig}$) are determined from signal MC samples convolved with a 2D Gaussian function,
which accounts for the difference in mass resolution between data and MC simulation. 
This 2D Gaussian function is extracted by using two Gaussian functions to fit the 2D $M(n)$ and $M(\pip\pim)$ spectra
in data with the correlation between them taken into account.

In the fit, the probability density functions of the signal and sideband regions are constructed as:
 \begin{equation}
  \label{fit1}
 \begin{aligned}
f_{\rm signal}=&N_{\rm sig}\cdot f_{\rm sig}+N_{\rm \Lambda_{c}^{+}bkg}\cdot f_{\rm \Lambda_{c}^{+}bkg}\\
&+N_{\operatorname{\rm non-\Lambda_{c}^{+}}}\cdot f_{\operatorname{\rm non-\Lambda_{c}^{+}}},\\
  \end{aligned}
 \end{equation}
 
  \begin{equation}
  \label{fit2}
 \begin{aligned}
f_{\rm sideband}&=N^{'}_{\operatorname{\rm non-\Lambda_{c}^{+}}}\cdot f_{\operatorname{\rm non-\Lambda_{c}^{+}}}\\
&=h\cdot N_{\operatorname{\rm non-\Lambda_{c}^{+}}}\cdot f_{\operatorname{\rm non-\Lambda_{c}^{+}}},\\
  \end{aligned}
 \end{equation}
where $N_{\rm sig}$, $N_{\rm \Lambda_{c}^{+} bkg}$, and $N_{\operatorname{\rm non-\Lambda_{c}^{+}}}$ denote the signal yield, the $\Lambda_c^+\bar{\Lambda}_c^-$ background yield, and the non$\operatorname{-}\Lambda_c^+$ background yield in the signal region, $N^{'}_{\rm non\operatorname{-}\Lambda_{c}^{+}}$ denotes the non$\operatorname{-}\Lambda_c^+$ yield in the sideband region. The ratio $h$ between $N^{'}_{\rm non\operatorname{-}\Lambda_{c}^{+}}$ and $N_{\rm non\operatorname{-}\Lambda_{c}^{+}}$ is fixed to 1.262 by fitting the $M_{\rm BC}$ distribution combined 11 ST modes and calculating the ratio between the number of background events in the sideband region and that of the signal region. This method ensures that the non$\operatorname{-}\Lambda_{c}^{+}$ (or $q\bar q$) background is well-estimated and as done in the previous BESIII measurement of $\Lambda_c^+\to nK_{S}^{0}\pi^+$\cite{BESIII:2016yrc}.
The background from the $\Lambda_{c}^{+}$ decays, $f_{\rm \Lambda_{c}^{+}bkg}$, is described by a shape extracted from the inclusive MC sample. The background from the non-$ \Lambda_{c}^{+}$ decay is modeled as: 
 \begin{equation}
 \label{hadron_pdf}
 \begin{aligned}
 f_{\rm non\operatorname{-}\Lambda_{c}^{+}}=\,((1-k)\cdot f_{1}+k\cdot f_{K_{S}^{0}}\,)\cdot f_{2}, 
 \end{aligned}
 \end{equation}
where $f_{1}$ and $f_{2}$ represent the first-order Chebychev polynomials, $k$ is the fraction of the $K_{S}^{0}$ 
component and free in the fit, and $ f_{K_{S}^{0}}$ represents the shape of the $\Ks$ signal extracted 
from the signal MC samples which has the same $\Ks$ shape as the background process.

Fig.~\ref{fig:fits} shows the fit results. 
The total signal yield of $\Lambda_c^+\to nK_{S}^{0}\pi^+\pi^0$ summing over eleven ST modes and seven c.m.~energies is determined
to be $N^{\rm DT} = 98 \pm 15$.  
The BF of $\Lambda_c^+\to nK_{S}^{0}\pi^+\pi^0$ is calculated by 
\begin{equation}
	\mathcal{B} = \frac{N^{\rm DT}}{\sum_{ij}N_{ij}^{\rm ST}\cdot(\varepsilon_{ij}^{\rm DT}/\varepsilon_{ij}^{\rm ST})\cdot\mathcal{B}_{\rm int}},
	\label{eq:br}
\end{equation}
where the indices $i$ and $j$ are the eleven ST modes and seven c.m.~energies, respectively, and  $\mathcal{B}_{\rm int}$ are the BFs of $\Ks\to\pip\pim$ and $\pi^0\to\gamma\gamma$~\cite{ParticleDataGroup:2020ssz}. 
The DT efficiencies $\varepsilon_{ij}^{\rm DT}$, ST yields $N^{\rm ST}_{ij}$, and ST efficiencies $\varepsilon_{ij}^{\rm ST}$ are listed in Tables~\ref{tab:N_ST}, \ref{tab:E_ST}, and \ref{tab:E_DT_1}, respectively.
The significance considering systematic uncertainty (see below) is evaluated to be $9.2 \sigma$ via $\sqrt{-2\times\Delta \ln {\mathcal L}}$, where $\Delta{\ln\mathcal L}$ is the variation in $\ln{\mathcal L}$ of the likelihood fit with and without the signal component included. 
Here, the method of considering systematic uncertainty is the same as the systematic uncertainty from fitting model described in the next section, and the minimum significance among all the fitting models is taken as the final significance.

\begin{table*}[htbp]
	\centering
	\caption{The ST yields, $N_{ij}^{\rm ST}$, at seven c.m.~energies. The uncertainties are statistical only.}
	\label{tab:N_ST}
	\bgroup
	\def\arraystretch{1.3}
    \begin{tabular}{l|cccccccc}
	   \hline\hline
	   Tag mode             & 4599.53 MeV    & 4611.86 MeV   & 4628.00 MeV   & 4640.91 MeV    & 4661.24 MeV   & 4681.92 MeV     & 4698.82 MeV \\
	   \hline
	   $\bar{p}\Ks$                & $1243\pm35$   & $~\,226\pm15$ & $~\,994\pm33$   & $1048\pm34$   & $1044\pm33$ & $~\,3141\pm57$   & $~\,889\pm30$   \\
	   $\bar{p}K^{+}\pim$          & $6607\pm89$   & $1094\pm37$ & $5513\pm37$  & $5842\pm83$   & $5447\pm79$ & $~\,15919\pm134$ & $4680\pm73$     \\
	   $\bar{p}\Ks\piz$            & $~\,587\pm33$ & $~\,119\pm16$ & $~\,569\pm33$   & $~\,552\pm33$    & $~\,527\pm32$  & $~\,1591\pm56$   & $~\,414\pm30$   \\
	   $\bar{p}\Ks\pim\pip$        & $~\,594\pm33$ & $~\,100\pm15$ & $~\,475\pm30$   & $~\,484\pm30$    & $~\,487\pm21$  & $~\,1365\pm51$   & $~\,414\pm28$   \\
	   $\bar{p}K^{+}\pim\piz$      & $1965\pm71$   & $~\,331\pm30$ & $1453\pm75$  & $1458\pm63$   & $1460\pm63$ & $~~~4361\pm109$  & $1172\pm62$     \\
	   $\bar{\Lambda}\pim$         & $~\,738\pm27$ & $~\,116\pm11$ & $~\,636\pm27$   & $~\,664\pm27$    & $~\,624\pm26$  & $~\,1916\pm45$   & $~\,495\pm23$   \\
	   $\bar{\Lambda}\pim\piz$     & $1681\pm54$   & $~\,281\pm22$ & $1342\pm50$  & $1483\pm50$   & $1338\pm46$ & $~\,3900\pm78$   & $1145\pm43$     \\
	   $\bar{\Lambda}\pim\pip\pim$ & $~\,744\pm35$ & $~\,130\pm14$ & $~\,547\pm31$   & $~\,690\pm34$    & $~\,703\pm33$  & $~\,1847\pm55$   & $~\,569\pm31$   \\
	   $\bar{\Sigma}^{0}\pim$      & $~\,502\pm25$ & $~~~95\pm12$ & $~\,384\pm22$   & $~\,413\pm23$    & $~\,414\pm22$  & $~\,1267\pm38$   & $~\,334\pm20$   \\
	   $\bar{\Sigma}^{-}\piz$      & $~\,309\pm24$ & $~~~68\pm10$ & $~\,242\pm21$   & $~\,271\pm22$    & $~\,264\pm22$  & $~\,~\,770\pm38$    & $~\,216\pm21$\\
	   $\bar{\Sigma}^{-}\pim\pip$  & $1146\pm47$   & $~\,204\pm21$ & $~\,922\pm19$   & $~\,995\pm46$    & $~\,949\pm44$  & $~\,2729\pm79$   & $~\,848\pm42$   \\
	   \hline\hline
   	\end{tabular}
   	\egroup
\end{table*}

\begin{table*}[htbp]
	\centering
	\caption{The ST efficiencies, $\varepsilon_{ij}^{\rm ST}$\%, at seven c.m.~energies. The uncertainties are statistical only. The quoted efficiencies do not include any sub-decay BFs.}
	\label{tab:E_ST}
	\bgroup
	\def\arraystretch{1.3}
    \begin{tabular}{l|cccccccc}
	   \hline\hline
	   Tag mode & 4599.53 MeV    & 4611.86 MeV   & 4628.00 MeV   & 4640.91 MeV    & 4661.24 MeV   & 4681.92 MeV     & 4698.82 MeV \\
	   \hline
	   $\bar{p}\Ks$                    & $54.6\pm0.2$ & $50.8\pm0.6$ & $48.9\pm0.2$ & $47.9\pm0.2$ & $46.4\pm0.2$ & $45.2\pm0.1$ &  $44.1\pm0.2$  \\
	   $\bar{p}K^{+}\pim$              & $49.9\pm0.1$ & $47.8\pm0.2$ & $46.1\pm0.1$ & $45.3\pm0.1$ & $44.3\pm0.1$ & $42.8\pm0.1$ &  $41.9\pm0.1$  \\
	   $\bar{p}\Ks\piz$                & $22.2\pm0.2$ & $20.8\pm0.4$ & $19.2\pm0.2$ & $19.1\pm0.2$ & $18.2\pm0.2$ & $17.6\pm0.1$ &  $16.7\pm0.2$  \\
	   $\bar{p}\Ks\pim\pip$            & $22.8\pm0.2$ & $20.4\pm0.4$ & $19.2\pm0.2$ & $19.3\pm0.2$ & $18.3\pm0.2$ & $18.7\pm0.1$ &  $17.4\pm0.2$  \\
	   $\bar{p}K^{+}\pim\piz$          & $19.4\pm0.1$ & $18.1\pm0.2$ & $16.8\pm0.1$ & $16.2\pm0.1$ & $ 15.7\pm0.1$& $15.4\pm0.0$ &  $14.9\pm0.1$  \\
	   $\bar{\Lambda}\pim$             & $47.1\pm0.3$ & $44.2\pm0.6$ & $40.7\pm0.3$ & $40.2\pm0.3$ & $38.8\pm0.3$ & $38.2\pm0.2$ &  $36.2\pm0.3$  \\
	   $\bar{\Lambda}\pim\piz$         & $20.8\pm0.1$ & $18.4\pm0.2$ & $17.6\pm0.1$ & $17.5\pm0.1$ & $ 16.9\pm0.1$& $16.1\pm0.1$ &  $15.7\pm0.1$  \\
	   $\bar{\Lambda}\pim\pip\pim$     & $15.1\pm0.1$ & $12.7\pm0.3$ & $12.7\pm0.1$ & $13.2\pm0.1$ & $12.7\pm0.1$ & $12.5\pm0.1$ &  $13.0\pm0.1$  \\
	   $\bar{\Sigma}^{0}\pim$          & $28.4\pm0.2$ & $24.8\pm0.5$ & $25.3\pm0.2$ & $24.2\pm0.2$ & $24.0\pm0.2$ & $23.2\pm0.1$ &  $21.9\pm0.2$  \\
	   $\bar{\Sigma}^{-}\piz$          & $22.8\pm0.3$ & $21.0\pm0.6$ & $21.5\pm0.3$ & $22.3\pm0.3$ & $20.5\pm0.3$ & $19.6\pm0.1$ &  $18.3\pm0.3$  \\
	   $\bar{\Sigma}^{-}\pim\pip$      & $24.5\pm0.1$ & $23.8\pm0.3$ & $21.9\pm0.1$ & $21.6\pm0.1$ & $20.9\pm0.1$ & $20.0\pm0.1$ &  $19.9\pm0.1$  \\
	   \hline\hline
   	\end{tabular}
   	\egroup
\end{table*}

\begin{table*}[!htbp]
	\centering	
	\caption{The DT efficiencies of $\Lambda_c^+\to nK_{S}^{0}\pi^+\pi^0$, $\varepsilon_{ij}^{\rm DT}$\%, at seven c.m.~energies. The uncertainties are statistical only. The quoted efficiencies do not include any sub-decay BFs.}
	\label{tab:E_DT_1}
	\bgroup
	\def\arraystretch{1.3}
    \begin{tabular}{l|cccccccc}
	   \hline\hline
	   Tag mode & 4599.53 MeV    & 4611.86 MeV   & 4628.00 MeV   & 4640.91 MeV    & 4661.24 MeV   & 4681.92 MeV     & 4698.82 MeV \\
	   \hline
	   $\bar{p}\Ks$                    & $8.3\pm0.1$ & $8.0\pm0.1$ & $7.5\pm0.1$& $7.0\pm0.1$& $6.8\pm0.1$& $6.7\pm0.1$ &  $6.5\pm0.1$ \\
	   $\bar{p}K^{+}\pim$              & $8.1\pm0.1$ & $7.3\pm0.1$ & $7.1\pm0.1$& $7.0\pm0.1$& $7.0\pm0.1$& $6.6\pm0.2$ &   $6.4\pm0.1$ \\
	   $\bar{p}\Ks\piz$                & $3.1\pm0.1$  & $2.9\pm0.1$  & $2.8\pm0.1$ & $2.6\pm0.1$ & $2.7\pm0.1$ & $2.6\pm0.1$  &  $2.4\pm0.1$  \\
	   $\bar{p}\Ks\pim\pip$            & $3.0\pm0.1$  & $2.5\pm0.1$  & $2.4\pm0.1$ & $2.4\pm0.1$ & $2.4\pm0.1$ & $2.3\pm0.1$  &  $2.3\pm0.1$  \\
	   $\bar{p}K^{+}\pim\piz$          & $3.1\pm0.1$  & $2.8\pm0.1$  & $2.6\pm0.1$ & $2.6\pm0.1$ & $2.4\pm0.1$ & $2.3\pm0.1$  &   $2.3\pm0.1$  \\
	   $\bar{\Lambda}\pim$             & $7.2\pm0.1$ & $6.3\pm0.1$ & $5.9\pm0.1$& $5.9\pm0.1$& $5.9\pm0.1$& $5.5\pm0.2$ &  $5.5\pm0.1$ \\
	   $\bar{\Lambda}\pim\piz$         & $3.0\pm0.1$  & $2.7\pm0.1$  & $2.5\pm0.1$ & $2.3\pm0.1$ & $2.3\pm0.1$ & $2.3\pm0.1$  &   $2.3\pm0.1$  \\
	   $\bar{\Lambda}\pim\pip\pim$     & $2.0\pm0.1$  & $1.8\pm0.1$  & $1.7\pm0.1$ & $1.7\pm0.1$ & $1.7\pm0.1$ & $1.7\pm0.1$  &  $1.7\pm0.1$  \\
	   $\bar{\Sigma}^{0}\pim$          & $4.3\pm0.1$ & $3.8\pm0.1$ & $3.6\pm0.1$ & $3.5\pm0.1$ & $3.5\pm0.1$ & $3.3\pm0.1$  &  $3.3\pm0.1$  \\
	   $\bar{\Sigma}^{-}\piz$          & $3.5\pm0.1$  & $3.4\pm0.1$  & $3.2\pm0.1$ & $3.1\pm0.1$ & $2.8\pm0.1$ & $2.7\pm0.1$  &  $2.6\pm0.1$  \\
	   $\bar{\Sigma}^{-}\pim\pip$      & $3.8\pm0.1$ & $3.6\pm0.1$  & $3.4\pm0.1$ & $3.3\pm0.1$ & $3.2\pm0.1$ & $3.0\pm0.1$  &  $2.9\pm0.1$  \\
	   \hline\hline
   	\end{tabular}
   	\egroup
\end{table*}

In order to consider potential intermediate resonant states in the decay of $\Lambda_c^+\to nK_{S}^{0}\pi^+\pi^0$, the distributions of the momenta of the four daughter particles $p(K_S)$, $p(n)$, $p(\pi^0)$, $p(\pi^+)$, and the invariant mass $M(n\pi^+\pi^0)$ are re-weighted according to 
the data to obtain the DT efficiency.
The derivations of the ST yields and ST efficiencies are the same as Ref.~\cite{BESIII:yinghao_work}.  
The BF of $\Lambda_c^+\to nK_{S}^{0}\pi^+\pi^0$ is calculated to be $(0.85\pm0.13)\%$, where the uncertainty is statistical only.

\section{\boldmath Systematic uncertainties}

The systematic uncertainties include the $\pi^+$ tracking and PID, $\pi^0$ and $K_{S}^{0}$ reconstruction, fitting models of tag and signal sides, MC statistics, mass window of peaking backgrounds, and MC model, as summarized in Table~\ref{tab:syst summary}.

The systematic uncertainty from the $\pi^+$ tracking and PID is studied by using the control sample $e^+e^-\to K^+K^-\pi^+\pi^-$.
An alternative efficiency is calculated by re-weighting events with momentum-dependent efficiency correction factors
extracted from the control sample.  
The difference between the nominal and alternative efficiencies, 0.3\%, is taken as systematic uncertainty.

The systematic uncertainties due to the reconstruction of $\pi^0$ and $K_{S}^{0}$ candidates are determined using the control samples $e^+e^-\to K^+K^-\pi^+\pi^-\pi^0$ and $J/\psi\to K^{*}(892)^{+}K^-\to K_{S}^{0}\pi^+K^-$. The difference of efficiencies between data and MC simulation is estimated with the same method used for $\pip$ tracking and PID. The systematic uncertainty is estimated to be 0.9\% for $\Ks$ reconstruction and 0.2\% for $\pi^0$ reconstruction.
The uncertainties of the BFs of $\pi^0\to\gamma\gamma$ and $K_{S}^{0}\to\pi^+\pi^-$, which are quoted from the PDG~\cite{ParticleDataGroup:2020ssz}, are 0.03\% and 0.1\%, respectively.

Uncertainties on the ST yields, DT efficiencies, and ST efficiencies all contribute to systematic uncertainties. 
The propagated uncertainties, as described by Eq.~(\ref{eq:br}), lead to a total uncertainty of 0.4\%. 

The uncertainty of 0.2\% on the fitting model of the ST side is quoted from Ref.~\cite{BESIII:yinghao_work}. 
To estimate the uncertainty of the 2D signal shape in the fit, 
we use two Gaussian functions to describe the signal contribution and take the difference of the fitted signal yields, 0.6\%,
as the systematic uncertainty.  
In the nominal fit, the ratio between $N^{'}_{\rm non\operatorname{-}\Lambda_{c}^{+}}$ and $N_{\rm non\operatorname{-}\Lambda_{c}^{+}}$, is fixed at $1.262\pm0.005$. To evaluate the systematic uncertainty from the ratio, it is varied by one standard deviation and the change of the signal yield is less than 0.1\%, which 
is negligible.
In the nominal fit, the background shape is exacted from the inclusive MC sample by RooKeysPdf~~\cite{Cranmer:2000du}. The uncertainty of the shape of $\Lambda_c^{+}$ background is considered by changing the smoothness factor of the RooKeysPdf from 0 to 1, and the difference of the signal yield, 1.8\%, is taken as the systematic uncertainty.  
Hence, the total systematic uncertainty associated with the fitting model of signal side is 1.9\%.

The systematic uncertainty of the peaking background window is estimated by using the control samples of $\Lambda_{c}^{+} \to \Sigma^{+}\,(n\pi^+) \, \pi^+\pi^-$, $\Sigma^{-}\,(n\pi^-) \, \pi^+\pi^+$, and $\Lambda\,(n\pi^0) \, \pi^+$. Gaussian functions are used to describe the difference between data and MC simulation. Using the simulated shapes convolved with Gaussian functions to fit the distributions of $M(n\pi^+)$, $M(n\pi^-)$, and $M(n\pi^0)$, the parameters of the Gaussian functions are determined.
The signal MC samples are smeared based on the widths and means of these Gaussian functions. The difference of the fitted signal yield, 0.3\%, is taken as the systematic uncertainty. 

The nominal DT efficiencies are calculated by weighting the signal MC samples. The efficiency difference between the weighted and PHSP signal MC samples, 3.0\%, is taken as the systematic uncertainty.

The total systematic uncertainty, 3.7\%, is obtained via the quadrature sum of the individual components.

\begin{table}[!h]
	\caption{Systematic uncertainties in the BF measurement. \label{tab:syst summary}}
	\begin{tabular}{cc}
	\hline\hline
	Source
		& $\Lambda_{c}^{+}\to n\Ks\pip\pi^0$ (\%)  \\ 
	\hline
	$\pi^+$ tracking and PID 
		& 0.3  \\
	$\pi^{0}$ reconstruction
		& 0.2 \\
	$K_{S}^{0}$ reconstruction
		& 0.9 \\	
	${\mathcal B}(K_{S}^{0}\to\pi^+\pi^-)$
		& 0.1 \\
	MC statistics
		& 0.4  \\
	Fitting model of tag side
		& 0.2  \\
	Fitting model of signal side 
		& 1.9  \\
	{Peaking background window}
		& 0.3 \\
	{MC model}
		&3.0 \\
	\hline
	Total
		&3.7\\
	\hline\hline
	\end{tabular}
\end{table}

\section{\boldmath Summary}

In summary, based on 4.5 fb$^{-1}$ of $\ee$ collision data samples taken at c.m.~energies between $4599.53\,\mev$ and $4698.82\,\mev$ with the BESIII detector, the BF of $\Lambda_c^{+}\to n K_{S}^{0} \pi^+\pi^0$ is measured to be $(0.85 \pm 0.13\pm 0.03)$\%, with a significance of 9.2$\sigma$, where the first uncertainty is statistical and the second systematic. 
Table~\ref{tab:br_fianl2} summarizes the BFs of $\Lambda_c^{+}\to n\Ks\pi^+\pi^0$ and its isospin partners.
The measured BF differs with the prediction of isospin statistical model~\cite{Gronau:2018vei}, $(1.54 \pm 0.08)$\% by $4.4\sigma$. This indicates that there may be resonant contributions or some unknown dynamics in this decay.
The total BF of the four body decay $\Lambda_c^{+}\to N \bar{K}\pi\pi$ is predicted to be $(12.88 \pm 0.69)$\%~\cite{Gronau:2018vei}. Our BF, together with the BFs of its isospin partners, offer important constraint on the theoretical prediction.

\begin{table}[htbp]
\centering
\footnotesize
\caption{The BFs of $\Lambda_c^{+}\to n K_{S}^{0} \pi^+\pi^0$ and its isospin partners.} 
\label{tab:br_fianl2}
\begin{tabular}{cc}
\hline \hline 
Decay mode& BF $(\times10^{-2})$\\\hline 
$\Lambda_c^+\to pK^-\pi^+\pi^0$ &$4.53 \pm 0.38$~\cite{BESIII:2015bjk} \\
$\Lambda_c^+\to nK^-\pi^+\pi^+$ &$1.90 \pm 0.12 $~\cite{BESIII:yinghao_work}\\ 
$\Lambda_c^+\to pK_{S}^{0}\pi^+\pi^-$ &$1.53 \pm 0.14$~\cite{BESIII:2015bjk}\\\hline
$\Lambda_c^{+}\to n K_{S}^{0} \pi^+\pi^0$ (Theoretical prediction)&$1.54 \pm 0.08$~\cite{Gronau:2018vei}\\ 
$\Lambda_c^{+}\to n K_{S}^{0} \pi^+\pi^0$ (This work)&$0.85 \pm 0.13\pm 0.03$ \\
 \hline \hline
\end{tabular}
\end{table}

\section{\boldmath ACKNOWLEDGMENTS}
The BESIII Collaboration thanks the staff of BEPCII and the IHEP computing center for their strong support. This work is supported in part by National Key R\&D Program of China under Contracts Nos. 2020YFA0406400, 2020YFA0406300; National Natural Science Foundation of China (NSFC) under Contracts Nos. 12375070, 11635010, 11735014, 11835012, 11935015, 11935016, 11935018, 11961141012, 12025502, 12035009, 12035013, 12061131003, 12192260, 12192261, 12192262, 12192263, 12192264, 12192265, 12221005, 12225509, 12235017; the Chinese Academy of Sciences (CAS) Large-Scale Scientific Facility Program; the CAS Center for Excellence in Particle Physics (CCEPP); Joint Large-Scale Scientific Facility Funds of the NSFC and CAS under Contract No. U2032108, No. U1832207; CAS Key Research Program of Frontier Sciences under Contracts Nos. QYZDJ-SSW-SLH003, QYZDJ-SSW-SLH040; 100 Talents Program of CAS; Fundamental Research Funds for the Central Universities, Lanzhou University, University of Chinese Academy of Sciences; The Institute of Nuclear and Particle Physics (INPAC) and Shanghai Key Laboratory for Particle Physics and Cosmology; European Union's Horizon 2020 research and innovation programme under Marie Sklodowska-Curie grant agreement under Contract No. 894790; German Research Foundation DFG under Contracts Nos. 455635585, Collaborative Research Center CRC 1044, FOR5327, GRK 2149; Istituto Nazionale di Fisica Nucleare, Italy; Ministry of Development of Turkey under Contract No. DPT2006K-120470; National Research Foundation of Korea under Contract No. NRF-2022R1A2C1092335; National Science and Technology fund of Mongolia; National Science Research and Innovation Fund (NSRF) via the Program Management Unit for Human Resources \& Institutional Development, Research and Innovation of Thailand under Contract No. B16F640076; Polish National Science Centre under Contract No. 2019/35/O/ST2/02907; The Swedish Research Council; U. S. Department of Energy under Contract No. DE-FG02-05ER41374.


\end{document}